\documentclass[11pt]{article}
\usepackage{epsfig,here,hangcaption,rotating,fancyheadings,amssymb,amsmath}
\input epsf
\catcode`\@=11
\def\marginnote#1{}

\def\draftlabel#1{{\@bsphack\if@filesw {\let\thepage\relax
  \xdef\@gtempa{\write\@auxout{\string
    \newlabel{#1}{{\@currentlabel}{\thepage}}}}}\@gtempa
    \if@nobreak \ifvmode\nobreak\fi\fi\fi\@esphack}
     \gdef\@eqnlabel{#1}}
\def\@eqnlabel{}
\def\@vacuum{}
\def\draftmarginnote#1{\marginpar{\raggedright\scriptsize\tt#1}}
\def\draft{\oddsidemargin -.5truein
        \def\@oddfoot{\sl preliminary draft \hfil
        \rm\thepage\hfil\sl\today\quad\militarytime}
        \let\@evenfoot\@oddfoot \overfullrule 3pt
        \let\label=\draftlabel
        \let\marginnote=\draftmarginnote

\def\@eqnnum{(\theequation)\rlap{\kern\marginparsep\tt\@eqnlabel}%
\global\let\@eqnlabel\@vacuum}  }
\def\preprint{\twocolumn\sloppy\flushbottom\parindent 1em
        \leftmargini 2em\leftmarginv .5em\leftmarginvi .5em
        \oddsidemargin -.5in    \evensidemargin -.5in
        \columnsep 15mm \footheight 0pt
        \textwidth 250mmin      \topmargin  -.4in
        \headheight 12pt \topskip .4in
        \textheight 175mm
        \footskip 0pt

\def\@oddhead{\thepage\hfil\addtocounter{page}{1}\thepage}
        \let\@evenhead\@oddhead \def\@oddfoot{} \def\@evenfoot{}
}
\def\titlepage{\@restonecolfalse\if@twocolumn\@restonecoltrue\o
necolumn
     \else \newpage \fi \thispagestyle{empty}\c@page\z@
        \def\thefootnote{\fnsymbol{footnote}} }
\def\endtitlepage{\if@restonecol\twocolumn \else  \fi
        \def\thefootnote{\arabic{footnote}}
        \setcounter{footnote}{0}}  
\catcode`@=12
\relax
\newcommand{\newc}{\newcommand}
\newc{\kevee}{keV$_{\mathrm{ee}}$ }
\newc{\dg}{$^{\mathrm{o}}$}

\begin{document}
\topmargin-1.cm
%
\begin{titlepage}
\vspace*{-64pt}

\begin{flushright}
{June 2001}\\
\end{flushright}

\vspace{1.8cm}

\begin{center}

\LARGE{\bf{Measurements of Scintillation Efficiency \\ and Pulse-Shape
for Low Energy Recoils \\ in Liquid Xenon.}}
\vspace*{1.3cm}

\Large{D. Akimov$^{\mathrm{a}}$, A. Bewick$^{\mathrm{b}}$,
D. Davidge$^{\mathrm{b}}$, J. Dawson$^{\mathrm{b}}$,\\
A.S. Howard$^{\mathrm{b}}$, I. Ivaniouchenkov$^{\mathrm{b}}$,
W.G. Jones$^{\mathrm{b}}$, M. Joshi$^{\mathrm{b}}$,\\
V.A. Kudryavtsev$^{\mathrm{c}}$, T.B. Lawson$^{\mathrm{c}}$,
V. Lebedenko$^{\mathrm{b}}$, M.J. Lehner$^{\mathrm{c}}$,\\
P.K. Lightfoot$^{\mathrm{c}}$, I. Liubarsky$^{\mathrm{b}}$,
R. L\"uscher$^{\mathrm{c}}$, J.E. McMillan$^{\mathrm{c}}$,\\
C.D. Peak$^{\mathrm{c}}$, J.J. Quenby$^{\mathrm{b}}$,
N.J.C. Spooner$^{\mathrm{c}}$, T.J. Sumner$^{\mathrm{b}}$,\\
D.R. Tovey$^{\mathrm{c}}$\footnote{e-mail: d.r.tovey@sheffield.ac.uk},
C.K. Ward$^{\mathrm{c}}$}\\
\vspace*{0.45cm}

\normalsize $^{\mathrm{a}}$ {\it Institute for Theoretical and
Experimental Physics, B. Cheremushkinskaja 25,\\ 117259 Moscow,
Russia.}

\normalsize $^{\mathrm{b}}$ {\it Physics Department, Blackett
Laboratory, Imperial College of Science, Technology \\ and Medicine,
Prince Consort Road, London SW7 2BZ, UK.}

\normalsize $^{\mathrm{c}}$ {\it Department of Physics and Astronomy,
University of Sheffield, Hounsfield Road, \\ Sheffield S3 7RH, UK.}

\end{center}

\bigskip

\begin{abstract}
Results of observations of low energy nuclear and electron recoil
events in liquid xenon scintillator detectors are given. The relative
scintillation efficiency for nuclear recoils is 0.22 $\pm$ 0.01 in the
recoil energy range 40 keV - 70 keV. Under the assumption of a single
dominant decay component to the scintillation pulse-shape the
log-normal mean parameter $T_0$ of the maximum likelihood estimator of
the decay time constant for 6 keV $<$ $E_{ee}$ $<$ 30 keV nuclear
recoil events is equal to 21.0 ns $\pm$ 0.5 ns. It is observed that
for electron recoils $T_0$ rises slowly with energy, having a value
$\sim$ 30 ns at $E_{ee}$ $\sim$ 15 keV. Electron and nuclear recoil
pulse-shapes are found to be well fitted by single exponential
functions although some evidence is found for a double exponential
form for the nuclear recoil pulse-shape.
\end{abstract}

\vspace{1cm}

\end{titlepage} 

\setcounter{footnote}{0}
\setcounter{page}{0}
\newpage

\section{Introduction}
Searches for Weakly Interacting Massive Particles (WIMPs) which may
constitute the galactic dark matter are currently being carried out by
a number of groups world-wide \cite{pfs1,cb1}. These experiments
generally seek evidence for anomalous populations of low energy ($<$
50 keV) nuclear recoil signals caused by elastic scattering of WIMPs
\cite{wg}. Efficient searches require targets which both couple
strongly to WIMPs, to give measurable event rates, and give a high
energy transfer. These requirements tend to favour heavy nuclei with
significant coherent spin-independent coupling and a nuclear mass
comparable to that of the most likely WIMP candidate, the SUSY
neutralino. In addition, a good experiment requires a low electron
equivalent energy threshold, as the predicted spectrum falls
exponentially with energy, and the ability to discriminate nuclear
recoil signals from the dominant background, which tends to be
electron recoil events.

Liquid xenon scintillator detectors fulfill all of these criteria. In
particular the quenching factor of xenon recoils, defined as the ratio
of the numbers of photons emitted from pure xenon by nuclear and
electron recoils of the same energy, is expected to be higher than
those of heavy nuclei in other scintillators \cite{gjd1}. Further, the
scintillation pulse-shapes of low energy nuclear recoils are expected
to be faster ($\lesssim$ 20 ns) than those of electron recoils of the
same energy and this permits some level of background rejection using
Pulse-Shape Analysis (PSA) \cite{rbo1}.

The nuclear recoil quenching factor cannot be measured directly.
Instead it must be approximated by the nuclear recoil relative
scintillation efficiency, defined as the ratio of the observed
scintillation pulse-height from a nuclear recoil to that from a
gamma-ray interaction depositing the same total amount of energy. The
scintillation pulse-height is commonly measured in `electron
equivalent' energy units by calibration with a mono-energetic
gamma-ray source producing electron recoils. The nuclear recoil
relative scintillation efficiency is then given by the ratio of the
measured electron equivalent energy $E_{ee}$ to the calculated nuclear
recoil energy $E_R$. In the limit where negligible recoil energy
(electron or nuclear) is lost through detector dependent secondary
processes (such as the absorption of ionisation by impurities, which
prevents emission of recombination photons) the quenching factor and
nuclear recoil relative scintillation efficiency are identical.

There has been some disagreement in the literature surrounding the
value of the nuclear recoil relative scintillation
efficiency/quenching factor in liquid xenon. Theoretical and
experimental work has suggested values ranging from $\simeq$ 0.2 up to
0.8 \cite{gjd1,rb1,fa1}. Given the relevance of this quantity to the
sensitivity of liquid xenon detectors to WIMP signal events it is
important that the true value is measured accurately. It is also
crucial that the scintillation pulse-shape properties of nuclear and
electron recoils in liquid xenon are determined in order to permit the
use of PSA with data from operational detectors such as that proposed
by the UK Dark Matter Collaboration (ZEPLIN-I \cite{pfs2}).

The response of dark matter detectors to nuclear recoils can be
determined through the elastic scattering of monoenergetic
neutrons. This technique has been used in previous work to measure
relative scintillation efficiencies and pulse-shapes in NaI(Tl) and
CaF$_2$(Eu) \cite{njcs1,gjd2,njcs2,drt1,vak1}. In this letter we
describe studies of nuclear recoils in liquid xenon using the same
technique.

\section{Detector Apparatus}
Data from two different liquid xenon detectors were used in this
study. The first detector (A) consisted of a cylindrical xenon volume
30 mm (L) $\times$ 30 mm (diam.) viewed from below through a single
Spectrasil A quartz window by an Electron Tubes 9829 QA bialkali
photomultiplier tube. High efficiency for collection of 175 nm xenon
scintillation photons was obtained by using a 5 mm thick PTFE
reflector lining the cylindrical wall of the chamber, together with a
PTFE float on the liquid surface. Stable light output of 0.90 $\pm$
0.02 photoelectrons per keV and energy resolution $\sigma(E)/E$ = 20.0
\% $\pm$ 0.7 \% was measured through calibration with the 122 keV line
from a 10 $\mu$Ci $^{57}$Co source. Calibration with the 662 keV line
from a 1 $\mu$Ci $^{137}$Cs source showed no evidence for significant
non-linearity. Further details regarding detector A can be found in
Ref.~\cite{tjs1}.

The second detector (B - designed by Chase Research Ltd.) consisted of
a xenon volume of dimension 78 mm (L) $\times$ 22 mm (diam.) defined
by a 27 mm thick annular PTFE reflector. The liquid xenon was viewed
by two ETL 9849 QB bialkali PMTs through Spectrasil A windows set at
the top and bottom of the chamber. Stable light output of 0.62 $\pm$
0.02 photoelectrons per keV and energy resolution $\sigma(E)/E$ = 22.9
\% $\pm$ 0.9 \% was measured using the same $^{57}$Co calibration
procedure as for detector A. Again no evidence for significant
non-linearity was found with higher energy sources.

Before use the xenon was purified using three processes: distillation,
passing through Oxisorb granules (supplied by Messer Griesheim GmbH)
and pumping out vapour present over the solid xenon formed at liquid
nitrogen temperature. Both detectors were baked and pumped out to a
pressure below 10$^{-7}$ Torr before filling with xenon. During tests
the liquid xenon was maintained at a temperature of -105 \dg C $\pm$
0.5 \dg C and a pressure of 1.3 bar (A) or 2 bar (B).

\section{Neutron Beam and Data Acquisition (DAQ) System}
The monoenergetic neutron beam used for the nuclear recoil
measurements was an Activation Technology Corporation d-d device at
the University of Sheffield, producing $\sim$ 10$^9$ neutrons/s
isotropically with energy 2.85 MeV. A double-wedged iron collimator of
mean diameter 19 mm, passing through borax, lead, iron and wax
shielding, was used to define the beam. Further details of the neutron
beam facility can be found in Ref.~\cite{drt1}.

For each test the liquid xenon detector was placed in the beam 0.2 m
from the collimator. A 75 mm diameter NE213 counter with pulse-shape
discrimination was placed at a distance of 400 mm from the target and
moved to a variety of angles $\theta$ to the beam direction. This was
surrounded by $\sim$ 200 mm of wax on all sides except that facing the
target detector so as to absorb wall-scattered neutrons. In this way
those events caused by the scattering of neutrons from xenon nuclei in
the target detector could be tagged with coincident nuclear recoil
events in the NE213 counter, and the xenon nuclear recoil energy
determined by kinematics from the neutron angle-of-scatter
$\theta$. Further details of the procedure can be found in
Refs.~\cite{njcs1,gjd2,njcs2}.

The data acquisition system used was similar to that described in
Ref.~\cite{drt1}. Pulses from the liquid xenon target detector went to
a discriminator with a trigger level set to accept single
photoelectrons. The signals from the discriminator were used to
provide start signals for a Time-to-Amplitude Convertor (TAC). A LINK
Systems 5020 pulse-shape discrimination unit was used to identify
nuclear recoil events in the NE213 counter and the output signals from
this were then used to stop the TAC. TAC pulses corresponding to time
differences less than 500 ns between the two signals triggered a
LeCroy Digital Sampling Oscilloscope (DSO), which digitised the event
pulse-shapes with 1 ns and 2 ns sampling times (for detectors A and B
respectively) with an 8 bit resolution. These data, together with the
TAC amplitude, were passed to a Macintosh PC running LabView DAQ
software for storage and off-line analysis.

\section{Data Analysis}
Data were taken at neutron scattering angles $\theta$ of 120\dg ,
105\dg and 90\dg . Peaked TAC amplitude distributions were observed
indicating low levels of background due to random coincidence between
gamma-ray induced events in the liquid xenon and the NE213 signals. In
subsequent software analysis data for study were selected using cuts
applied to the TAC amplitude distributions.

Scintillation events from the liquid xenon target detectors were
analysed by estimating the arrival times of photoelectrons in the
events. The procedure used involved integrating the event pulse-shapes
and then noting the times of equal pulse-height increment, using an
increment set at the mean expected pulse-height for a single
photoelectron. The mean photoelectron arrival time for each event
($\bar{t}$), calculated relative to the arrival time of the first
photoelectron in the event, was then used to estimate the event time
constant. Under the assumption that the pulse-shape is predominantly a
single exponential the $\bar{t}$ value provides the maximum likelihood
estimator of the decay time constant of the pulse \cite{drt1}.

\section{Relative Scintillation Efficiency}
\begin{figure}
\begin{center}
\epsfig{file=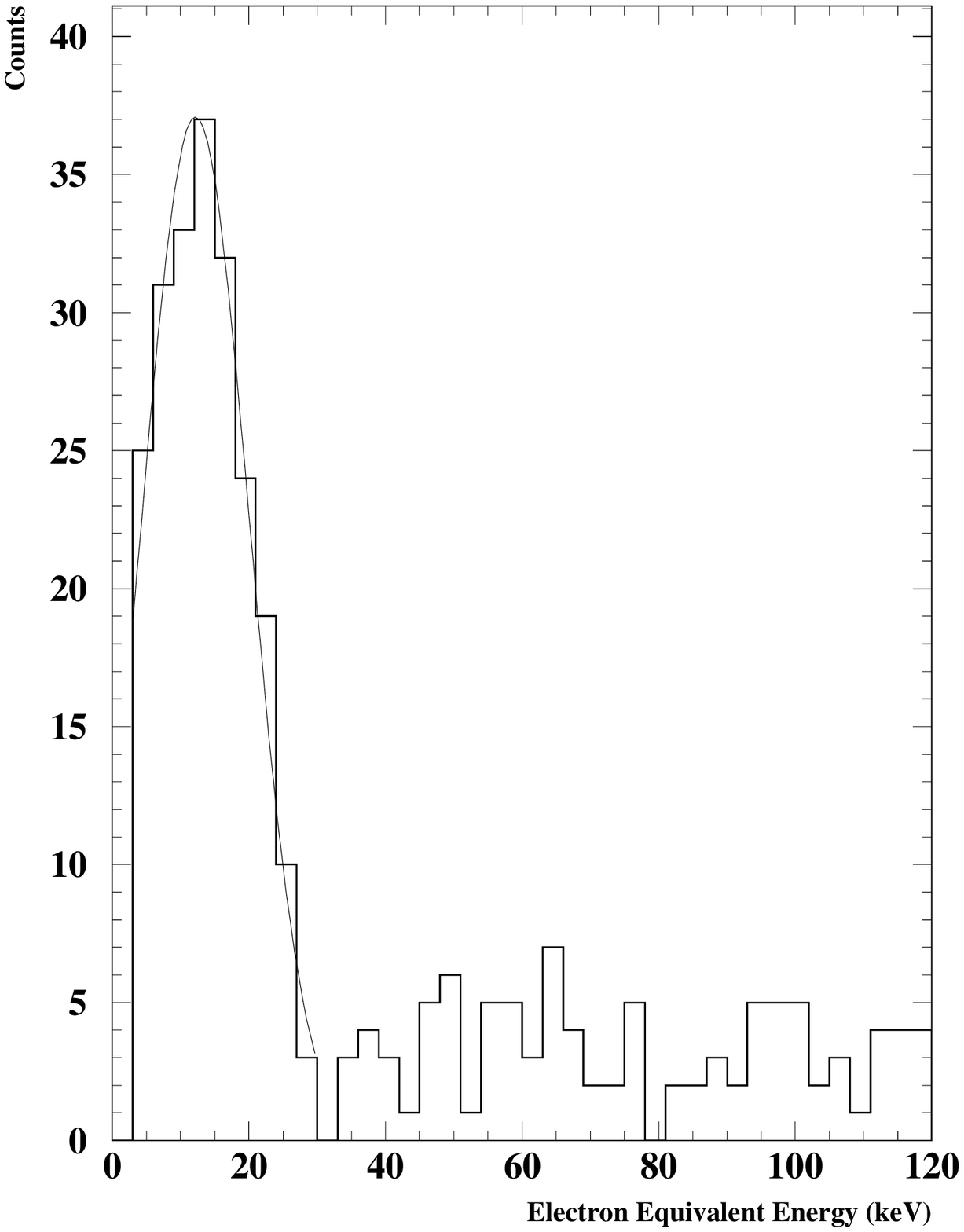,height=5.0in,}
\caption{\label{fig1}{\it}}
\end{center}
\end{figure}

Values of the relative scintillation efficiency for each scattering
angle are given by the ratio of the mean electron equivalent energy of
nuclear recoils and the expected nuclear recoil energy at that
angle. The mean electron equivalent energies of nuclear recoils were
determined by fitting gaussian curves to the observed energy
distributions of events (Fig.~\ref{fig1}). Only those events with
electron equivalent energies $<$ 30 keV were considered so as to avoid
contamination with high energy events from inelastic excitation and
decay of the 39.6 keV and 80.2 keV levels in $^{129}$Xe and $^{131}$Xe
. Cuts on the $\bar{t}$ values of events (12 ns $<$ $\bar{t}$ $<$ 30
ns) were also used to reduce this source of contamination. The
electron equivalent energy scale for the fits was determined from
energy calibrations with a $^{57}$Co source (122 keV line) according
to the calibration procedure described in \cite{drt1}. The
scintillation efficiency was thus measured with respect to this high
energy electron recoil calibration point.

\begin{figure}
\begin{center}
\epsfig{file=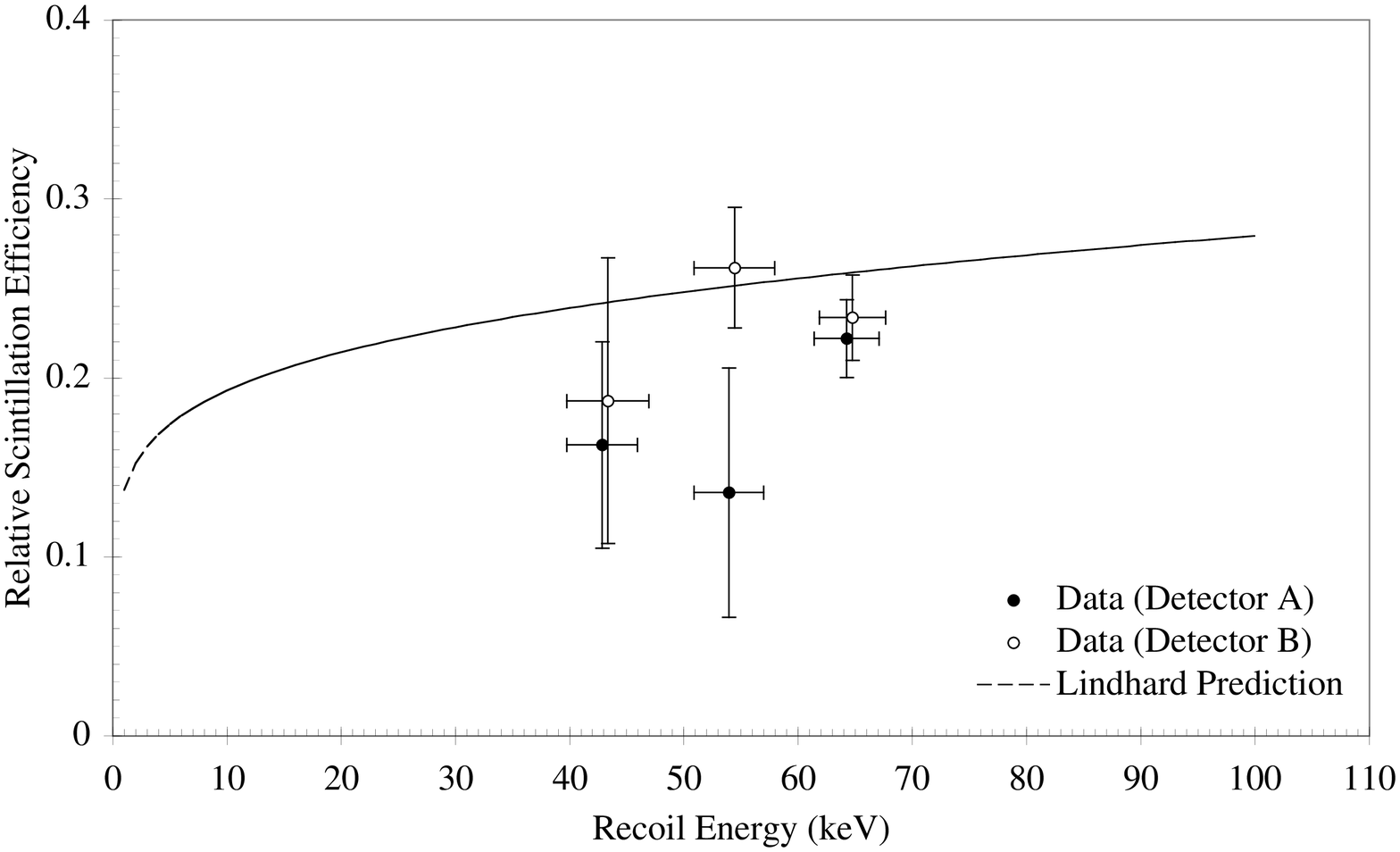,height=3.0in,}
\caption{\label{fig2}{\it}}
\end{center}
\end{figure}

Results of the relative scintillation efficiency measurements are
presented in Fig.~\ref{fig2} as a function of nuclear recoil energy
$E_R$. The mean values obtained from both detectors are consistent
within errors with a constant value of 0.22 $\pm$ 0.01 over the recoil
energy range 40 - 70 keV. The errors on the individual points include
both statistical errors determined from the fits and systematic
effects due to the finite solid angle acceptance of the NE213
counters, possible fluctuation in the neutron beam energy and
variation in fitted means with event selection parameters, histogram
binning etc.

The value of the liquid xenon relative scintillation efficiency
obtained from this work is in agreement with that obtained from recent
measurements carried out by the ICARUS collaboration \cite{fa1}, but
contrasts strongly with values reported by the DAMA collaboration
\cite{rb1}.

\section{Scintillation Time Constant}
Scintillation time constants for liquid xenon were estimated from the
distributions of event $\bar{t}$ values, which are approximately
log-normal for a single exponential pulse-shape \cite{drt2}. The
fitted log-normal mean-parameters \cite{pfs1} (here denoted by $T_0$ -
note that this is not the true mean of the log-normal) of these
distributions provide an estimate of the exponential time constant of
the parent scintillation pulse-shape \cite{drt2}.

\begin{figure}
\begin{center}
\epsfig{file=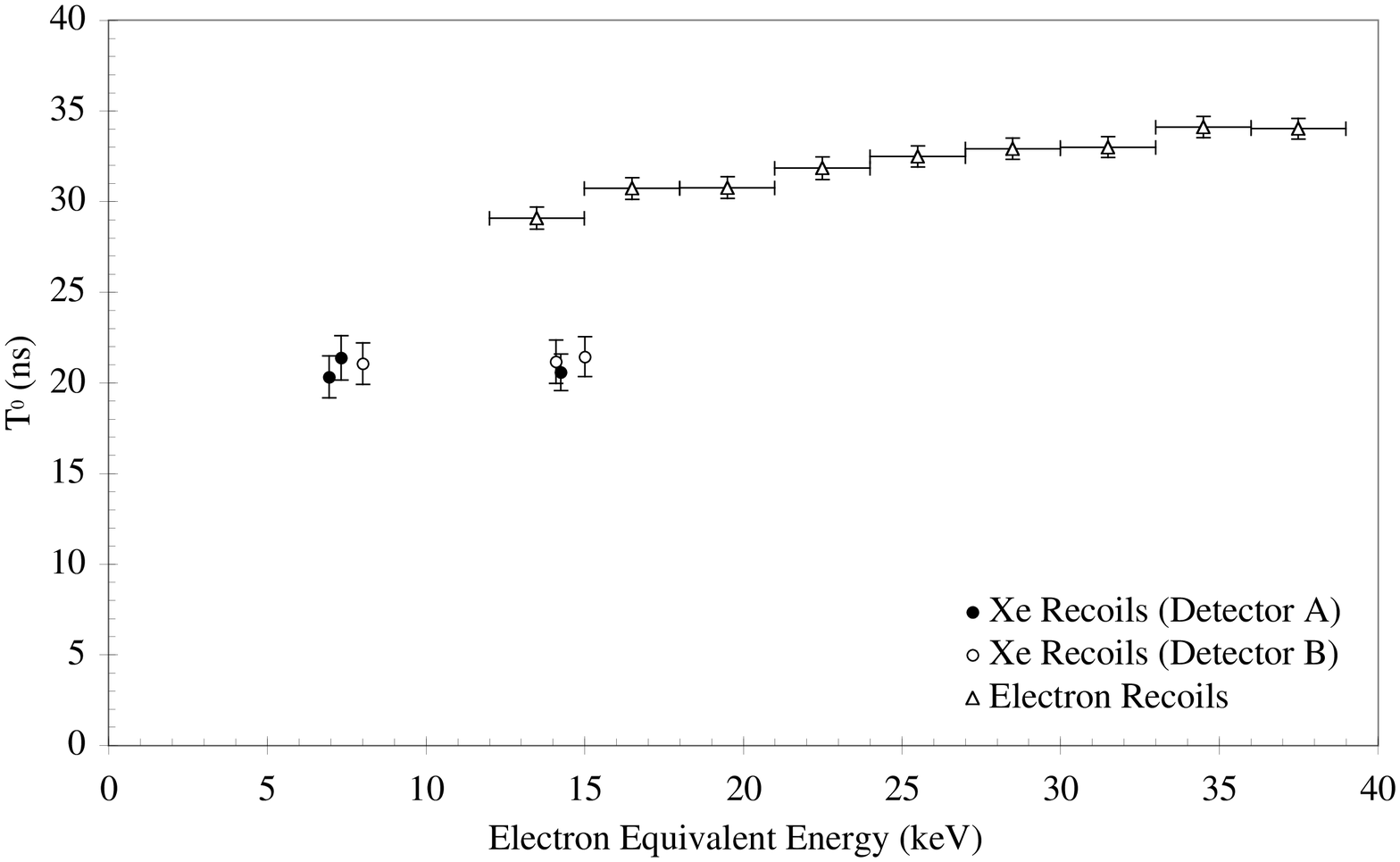,height=3.0in,}
\caption{\label{fig3}{\it}}
\end{center}
\end{figure}

$T_0$ values are plotted in Fig.~\ref{fig3} for both nuclear recoils
and electron recoils. Nuclear recoil $\bar{t}$ distributions for each
scattering angle were constructed from events in the range 6 keV $<$
$E_{ee}$ $<$ 30 keV (error bars not shown) but the plotted $E_{ee}$
values correspond to the fitted means of the $E_{ee}$ distributions
determined when measuring the relative scintillation
efficiency. Electron recoil $\bar{t}$ distributions were constructed
from pulses acquired during exposure of detector B to a 1 $\mu$Ci
$^{60}$Co source generating low energy Compton scattering events. This
data extends down to electron equivalent energies of 12 keV below
which the presence of significant numbers of fast ($\bar{t}$ $<$ 9 ns)
noise events (due to lack of an NE213 coincidence requirement)
prevents measurement of $T_0$. Note however that these events are not
present in the electron equivalent energy distributions used to
calculate relative scintillation efficiencies due to the application
of the $\bar{t}$ cut (12 ns $<$ $\bar{t}$ $<$ 30 ns). Vertical error
bars on all points include both statistical errors and a constant
systematic error (estimated to be $\sim$ 0.5 ns) due to uncertainties
in fitted values caused by event selection and binning.

The data show strong evidence for significant differences in $T_0$ and
hence scintillation time constant between nuclear and electron recoils
of the same electron equivalent energy. The $T_0$ values for nuclear
recoils from both detectors are consistent within errors with a
constant value of 21.0 ns $\pm$ 0.5 ns, while the values for electron
recoils vary from 29.1 ns $\pm$ 0.6 ns at $\sim$ 13.5 keV to 34.0 ns
$\pm$ 0.6 ns at $\sim$ 37.5 keV.

\section{Scintillation Pulse-Shape}
\begin{figure}
\begin{center}
\epsfig{file=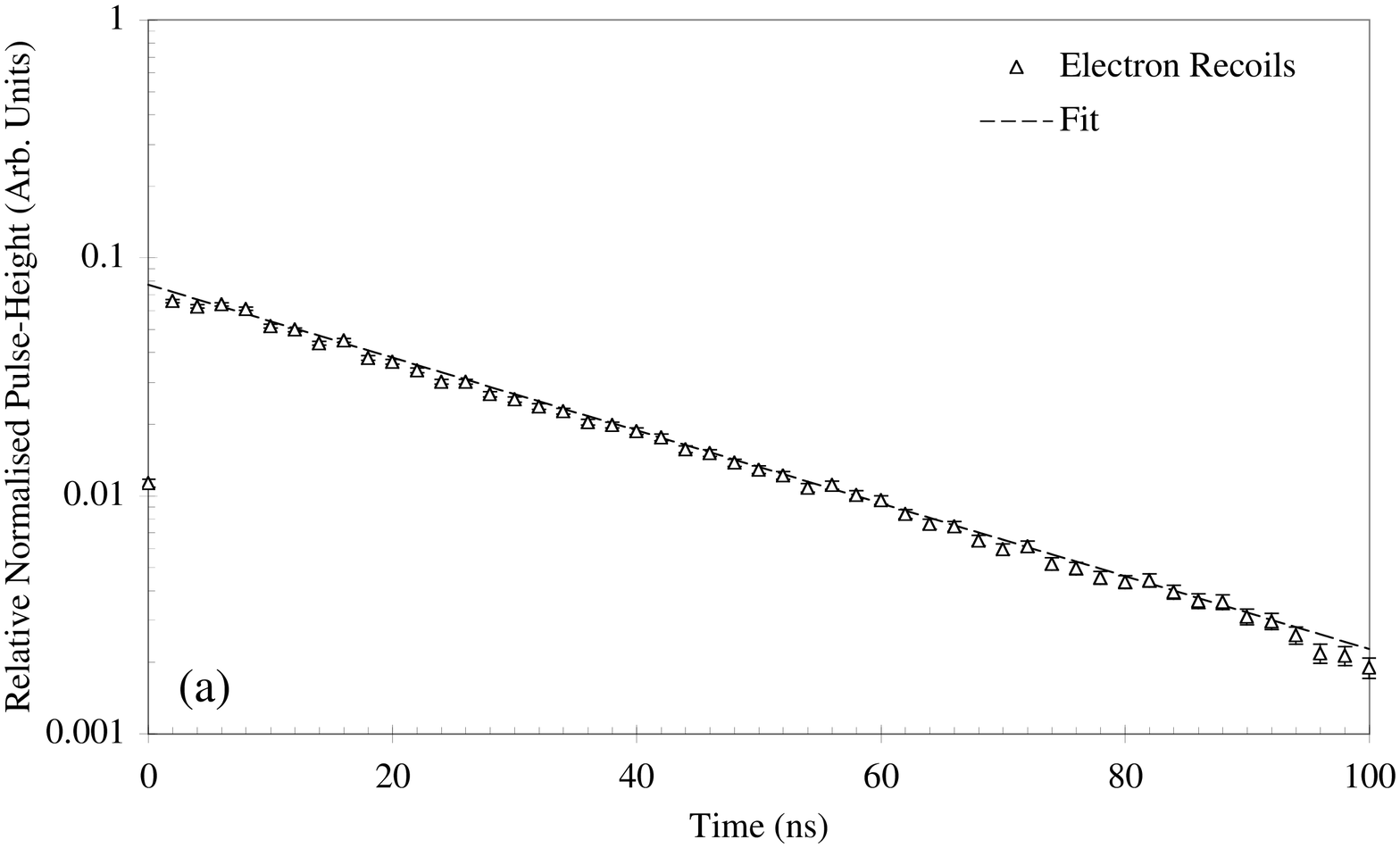,height=3.0in,}
\epsfig{file=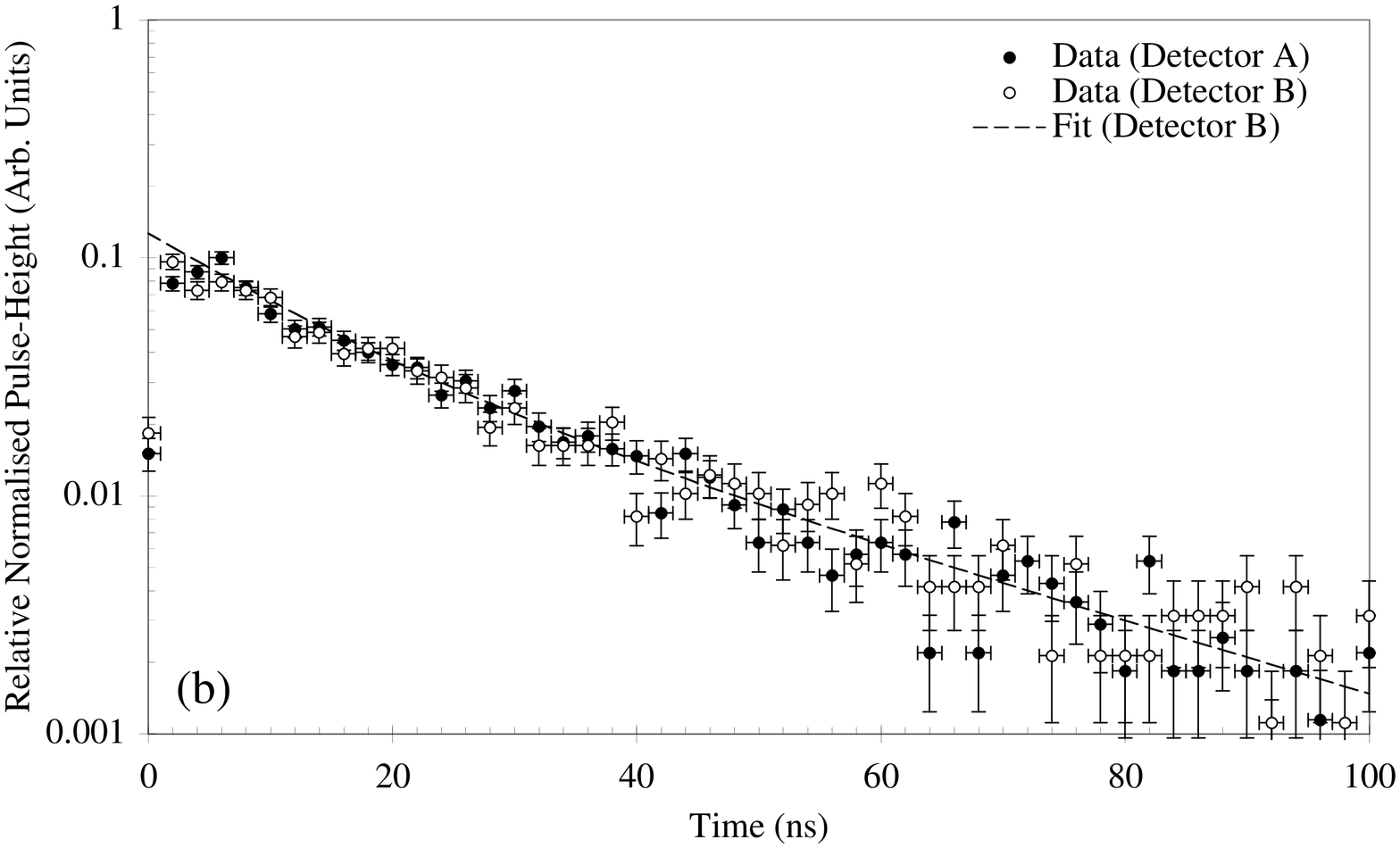,height=3.0in,}
\caption{\label{fig4}{\it}}
\end{center}
\end{figure}

In order to further investigate the scintillation pulse-shape of low
energy recoil events in liquid xenon normalised distributions were
constructed of the arrival times of photoelectrons relative to the
first photoelectron for events in the electron equivalent energy range
12 keV - 30 keV. These distributions are shown in Fig.~\ref{fig4} for
electron recoils (Fig.~\ref{fig4}(a)) and nuclear recoils
(Fig.~\ref{fig4}(b)). Due to poor event statistics in the latter case
distributions obtained from data taken at all three neutron scattering
angles have been summed. In both cases the data are well fitted by
single exponential functions, although the nuclear recoil data are
marginally better fitted by a sum of two exponentials
(e.g. $\chi^2$/dof = 1.117 for a double exponential fit to detector B
data with $\tau_1$ = 11.3 ns $\pm$ 4.1 ns and $\tau_2$ = 28.9 ns $\pm$
5.3 ns, as opposed to $\chi^2$/dof = 1.386 for a single exponential
fit with $\tau$ = 21.5 ns $\pm$ 0.76 ns). Such a double exponential
pulse-shape is expected on general grounds due to the action of at
least two different emission processes within the scintillator
\cite{gjd1}. It should be emphasised however that sensitivity to the
fastest components of such a pulse-shape is limited by the non-zero
transit-time jitter of the PMTs.

\section{Conclusions}
A 2.85 MeV mono-energetic neutron beam has been used to measure the
relative scintillation efficiency and the log-normal mean parameter
$T_0$ of the maximum likelihood time constant estimator for nuclear
recoils in liquid xenon. A mean value for the relative scintillation
efficiency of 0.22 $\pm$ 0.01 in the recoil energy range 40 keV - 70
keV was obtained. A $T_0$ value of 21.0 ns $\pm$ 0.5 ns was found for
nuclear recoil events with electron equivalent energy in the range 6
keV - 30 keV. Electron recoils were found to possess $T_0$ values
rising from 29.1 ns $\pm$ 0.6 ns at $\sim$ 13.5 keV to 34.0 ns $\pm$
0.6 ns at $\sim$ 37.5 keV thereby indicating significant pulse-shape
discrimination potential. Electron and nuclear recoil pulse-shapes
were found to be well fitted by single exponential functions although
some evidence was found for a double exponential form for the nuclear
recoil pulse-shape.

\section*{Acknowledgements}
The authors are grateful for discussions with P.F. Smith,
N.J.T. Smith, S. Chase and H. Wang. They would also like to thank
R. Nicholson and the staff of the University of Sheffield Central
Mechanical Workshop for their assistance. The authors wish to
acknowledge PPARC (DRT - Post-Doctoral Fellowship, RL - Royal Society
ESEP grant), Zinsser Analytic Ltd. (PKL, JEM), Electron Tubes
Ltd. (JEM, JWR) and Hilger Analytical Ltd. (JWR, CKW) for support.

\newpage

\section*{Figures}

Figure 1: Electron equivalent energy spectrum of events taken with
detector A at neutron scattering angle $\theta$ $=$ 120\dg \mbox{}
following TAC preselection only. The large peak at $E_{ee}$ $\simeq$
14 keV contains events with $\bar{t}$ $\sim$ 20 ns. Events in the
broad distribution at energies $\gtrsim$ 30 keV possess $\bar{t}$
$\gtrsim$ 30 ns characteristic of electron recoil events. These events
are consequently interpreted as being due to inelastic scattering of
neutrons. Elastic nuclear recoil peaks corresponding to relative
scintillation efficiencies of 0.45 and 0.65 (values quoted in
Ref.~\cite{rb1}) would lie at electron equivalent energies of 28.9 keV
and 41.8 keV respectively.
\\
\newline
Figure 2: Nuclear recoil relative scintillation efficiencies
$E_{ee}/E_R$ in liquid xenon as a function of nuclear recoil energy
$E_R$. Vertical error bars are dominated by statistical effects but
include both statistical and systematic contributions. Nuclear recoil
energies for points corresponding to data from detector B (open
circles) have been offset by 0.5 keV for clarity. The mean value is
0.22 $\pm$ 0.01. The dotted line corresponds to the expected value
from Lindhard theory \cite{jl1}.
\\
\newline
Figure 3: Fitted $T_0$ values for nuclear recoil and electron recoil
events as a function of electron equivalent energy. Vertical error
bars correspond to total errors obtained by adding statistical and
systematic errors in quadrature. Nuclear recoil $E_{ee}$ values are
those obtained when measuring the nuclear recoil relative
scintillation efficiency. Nuclear recoil points correspond to all
events in the electron equivalent energy range 6 keV $<$ $E_{ee}$ $<$
30 keV, although horizontal error bars are not shown for clarity.
\\
\newline
Figure 4: Photoelectron arrival time distributions for (a) electron
recoils (detector B) and (b) nuclear recoils (detectors A and B) in
the electron equivalent energy range 12 keV - 30 keV. Bin widths of 2
ns have been used and nuclear recoil data have been summed over all
three scattering angles. Vertical error bars show errors due to finite
photoelectron statistics. Also shown (dashed curves) are fits to a
single exponential function (Figure 4(a)) and a sum of two exponential
functions (Figure 4(b) - detector B data).
\newline

\newpage
\newpage
\newpage
\newpage

\end{document}